\documentclass[english,aps,prb,floatfix,twocolumn,showpacs]{revtex4}
\usepackage{amsmath}
\usepackage{babel}
\usepackage{graphicx}
\usepackage{amssymb}
\usepackage{multirow}

\begin{document}

\title{Antiferromagnetic metal phases in double perovskites having strong antisite defect concentrations}
\author{Prabuddha Sanyal} 
\affiliation{
IIT Roorkee, India} 
\pacs{}
\date{\today}

\begin{abstract}
Recently an antiferromagnetic metal phase has been proposed in 
double perovskites materials like Sr$_{2}$FeMoO$_{6}$ (SFMO), when electron doped. This material has been found to change from half-metallic ferromagnet to antiferromagnetic metal (AFM) upon La-overdoping. The original proposition of such an AFM phase was made for clean samples, but the experimental realization of La-overdoped SFMO has been found to contain a substantial fraction of antisite defects.
 A phase segregation into alternate Fe and Mo rich regions was observed.
In this paper we propose a possible scenario in which this type of strong antisite concentration can still 
result in an antiferromagnetic metal phase, by a novel hopping driven mechanism. Using variational calculations, we find that proliferation of such phase segregated domain regions can also result in the stabilization of an A-type AFM over the G-type AFM based on kinetic energy considerations. 
 
\end{abstract}
\maketitle
\noindent
\section{Introduction}

 Double perovskites, 
 with the general formula A$_{2}$BB$'$O$_{6}$, A=alkaline/rare-earth metals and B/B$'$=transition metals, are being studied extensively in recent years due to their half-metallic properties~\cite{tomioka,TSD}, and their importance in spintronic applications. A prominent member, Sr$_{2}$FeMoO$_{6}$, is ferromagnetic in the pure state with a $T_{c}$ 
as high as 410K.~\cite{SFMO,DDreview}. Substantial tunnelling magnetoresistance is also obtained in powdered samples of these compounds at low temperatures.
 ~\cite{SFMO,Mag-Res1,Mag-Res2,MRpaper}. There is a kind of defect which is possible in double perovskite materials called antisite defects in which a B-atom (Fe) and a B$^{'}$ atom (Mo) can interchange position, thereby bringing two B (Fe) atoms
 next to each other. Such defect regions tend to be antiferromagnetic and insulating, and decrease the overall
 magnetization. While  
 tunneling across grain boundaries, rather than antisite regions dominates as the 
main mechanism for magnetoresistance, the magnitude of the magnetoresistance does get affected quite dramatically by the 
 antisite defect concentration.~\cite{MRpaper}.
Attempts have been made to electron dope this compound (SFMO) hoping to enhance the $T_{c}$ and hence the polarization at room temperature and thereby improve the magnetoresistance.~\cite{JAP,ES,Navarro,photoemission}. While a 
slight increase in $T_{c}$ is indeed observed for low dopings~\cite{Navarro,p21n}, but for overdoping ($x>1$),
 the $T_{c}$ is found to decrease,as shown in~\cite{mePinaki,LSFMO}, and the ferromagnetism gets
replaced by antiferromagnetic phases. These antiferromagnetic phases are however, metallic, and are stabilized
 over the ferromagnetic phase due to a kinetic energy-driven mechanism. 
 This was derived using a model Hamiltonian in the clean limit without antisite defects, using analytical as well as numerical methods~\cite{mePinaki}. Further confirmation was obtained using abinitio methods, by  considering the Sr$_{2-x}$La$_{x}$FeMoO$_{6}$ series~\cite{LSFMO}, for the overdoped case ($x>1$). In the actual experimental preparation of this proposed series of materials, however, substantial concentration of antisite defects
 was found to be present~\cite{Sugatapriv}. In fact,
 a microscopic phase segregation situation was encountered, 
 where the sample phase segregated into alternate Fe-rich and
 Mo-rich regions, forming a patchy structure.  In this paper, using a variational approach, we reconsider the stability of magnetic phases in presence of antisite defects, 
 and in particular in the phase segregation scenario where alternate Fe-rich and Mo-rich regions are present. We
 find that antiferromagnetic metal phases are still stabilized kinetically for large regions of filling. The
  exact nature of the kinetically stabilized antiferromagnetic phase, however, changes from the clean limit
 to the alternate Fe-Mo rich phase segregated scenario. We confirm this change using numerical simulations based
 on exact diagonalization, where phase segregated Fe-Mo domains are allowed to proliferate over an otherwise clean
 sample. Our results are in consonance with those of an earlier study in presence of antisite defects
 using a different method, where it was found that antiferromagnetic phases do survive even in presence of 
 antisite disorder~\cite{VSinghPinaki}. Although they considered correlated antisite disorder, 
 they did not consider alternate Fe-Mo rich regions or phase segregation. We have obtained analytical results in
  the phase segregated limit, using a variational approach.
 The organization of the paper is as follows. In the next section, we shall briefly summarize the experimental results for the Sr$_{2-x}$La$_{x}$FeMoO$_{6}$ series (i.e., cataion site substitution), and the role of antisite defects. Next, we discuss a Hamiltonian suitable for analyzing this problem including antisite defects. In the next section, we consider the results. First we consider a fully phase segregated situation with alternate Fe-rich and Mo-rich segments, using analytical variational methods. Then we proceed to consider a generic situation where 
Fe-rich and Mo-rich domains are allowed to form inside an otherwise clean sample with rocksalt structure. Using
 exact diagomalization methods, we then consider the proliferation of such domains inside the sample.

\section{Experimental data on La-doped SFMO: Brief summary}

Experimental attempts at  electron doping  Sr$_{2}$FeMoO$_{6}$ by substituting Sr$^{2+}$  with La$^{3+}$ had been ongoing for a while~\cite{JAP,Navarro}. This results in the series of compounds 
 Sr$_{2-x}$La$_{x}$FeMoO$_{6}$. While some of these authors have reported
that the samples retained tetragonal symmetry (I4/mmm) upon electron doping from x=0 all the way upto x=1, others
report a change to monoclinic symmetry (P2$_{1}$/n) upon doping beyond x=0.4. However, all the authors consent
 on the fact that the cell parameters increase upon doping, and the antisite defect concentration increases, 
thereby reducing the degree and extent of ordering. This can be understood by considering the behaviour of the
saturation magnetization, and comparing with abinitio calculations. Such calculations indicate that the 
half-metallic behaviour persists from Sr$_{2}$FeMoO$_{6}$ 
 till SrLaFeMoO$_{6}$ (LSFMO)~\cite{JAP,LSFMO}, and the
 total moment changes from 4$\mu_{B}$/f.u. to 3$\mu_{B}/f.u.$, i.e., a reduction in moment by 1$\mu_{B}/f.u.$
 by doping by $x=1$, in the clean limit. This is due to the fact that a replacement of Sr$^{2+}$ ion by
 La$^{3+}$ ion due to La-doping results in the addition of an electron into the cell, 
and since LSFMO is half-metallic, the extra electron goes into
 the minority spin band, reducing the moment by exactly 1$\mu_{B}$.
 The experimental reduction, on the other hand, is by 2$\mu_{B}$/f.u.~\cite{JAP}, for example, clearly indicating the increase in concentration of antisite defects, along with the electron doping effect. Such increase in
antisite defect concentration has been confirmed in almost all the studies, even leading to a 
concentration level of $0.5$ in some cases for $x>0.8$~\cite{Navarro},
 where a phase separation scenario is possible. 

Recently, there has been a study of La-overdoping of Sr$_{2}$FeMoO$_{6}$, where the regime above $x>1$ is investigated~\cite{Sugatapriv}. As before, increase in antisite disorder with increasing doping is observed. However,
 the sample is found to phase segregate into alternate Sr,Mo-rich and La,Fe-rich short range patches, which is
 confirmed by X-ray spectra analysis. The number of Mo-O-Mo connections, which should be 0 for a perfectly ordered arrangement, 3 for a random situation, and 6 for a phase segregation scenario, was found to be more than 3, while the samples crystallized to a space group of P$2_{1}$/n, indicating a patchy structure as mentioned before. 
 Interestingly, even with this heavy concentration of antisite disorder, the samples remained metallic, although
 there was a transition from ferromagnetic to antiferromagnetic phase, as evidenced from magnetization hysteresis
 data. Thus, a rare antiferromagnetic metallic phase appears to be stabilized for an overdoped, phase-segregated
 sample of SFMO. In this paper we provide a plausible explanation to such a situation using model Hamiltonian 
 methods. While a crossover from ferromagnetic metal phase to an antiferromagnetic metal phase has already been
 predicted in the clean limit without antisites~\cite{mePinaki,LSFMO}, in this paper we consider the case with
 antisite defects, in particular a phase segregation scenario, and justify how this crossover can continue
 to hold even in this scenario.

\section{Hamiltonian}

A Hamiltonian suitable for considering general Fe-Mo configurations with antisite defects 
can be obtained in the following way~\cite{vivekanand,VSinghPinaki},
 by considering a local binary variable $\eta_{i}$ 
which is $1$ or $0$ depending on whether the site is Fe or Mo:

$$H= \epsilon_{Fe}\sum_{i,\sigma}\eta_{i}f_{i\sigma}^{\dagger}f_{i\sigma}+
\epsilon_{Mo}\sum_{i,\sigma}(1-\eta_{i})m_{i\sigma}^{\dagger}m_{i\sigma} $$

$$+t_{FF}\sum_{<ij>\sigma}\eta_{i}\eta_{j}f_{i\sigma}^{\dagger}f_{j\sigma} 
+t_{MM}\sum_{<ij>\sigma}(1-\eta_{i})(1-\eta_{j})m_{i\sigma}^{\dagger}m_{j\sigma} $$ 
$$+t_{FM}\sum_{<ij>\sigma}(\eta_{i}+\eta_{j}-2\eta_{i}\eta_{j})f_{i\sigma}^{\dagger}m_{j\sigma}$$ 
\begin{eqnarray}
+ J\sum_{i\alpha\beta}\eta_{i} {\bf S}_{i} \cdot
f_{i\alpha}^{\dagger}\vec{\sigma}_{\alpha\beta}f_{i\beta}+J_{AF}\sum_{<ij>}\eta_{i}\eta_{j}S_{i}\cdot S_{j}
\label{fullhamSFMO}
\end{eqnarray}

The $f$'s refer to the Fe sites and the $m$'s to the Mo sites.
$t_{FM}$, $t_{MM}$, $t_{FF}$ represent the nearest neighbor Fe-Mo, 
 Mo-Mo and Fe-Fe hoppings respectively.
 $\sigma$ is the spin index.  
 The difference between the ionic levels,
${\tilde \Delta} = \epsilon_{Fe} - \epsilon_{Mo}$, defines the charge transfer energy.
The ${\bf S}_i$ are `classical' (large $S$)
 core spins at the B site, coupled
to the itinerant B electrons through a coupling $J \gg t_{FM}$.  Each $\{\eta_{i}\}$ realization corresponds to a configuration of antisite disorder;
 in particular, the ordered configuration (without antisite defects) corresponds to a checkerboard,
 or $[\pi,\pi]$ configuration of $\eta_{i}$ on a 2D lattice. 
 The model then coincides with the
 two-sublattice Kondo lattice model which has been considered by several authors~\cite{Millis,Avignon,Guinea1,Guinea2}in the context of ordered double perovskites. If one considers the limit $J\rightarrow\infty$,
 then the model given by Eq~\ref{fullhamSFMO} simplifies to a model 
with ``spinless'' Fe degrees of freedom~\cite{Guinea1,Guinea2}:
\begin{eqnarray}
H &=& t_{FM}\sum_{<ij>\alpha}
     (\eta_{i}+\eta_{j}-2\eta_{i}\eta_{j})( sin(\frac{\theta_i}{2}) \tilde{f}_{i}^{\dagger}m_{j\uparrow} \cr
 &&-
e^{i\phi_i}cos(\frac{\theta_i}{2})\tilde{f}^{\dagger}_{i}m_{j\downarrow}) \cr
&& +h.c.
+t_{MM}\sum_{<ij>}(1-\eta_{i})(1-\eta_{j})m_{i\sigma}^{\dagger}m_{j\sigma}  \cr
&&
+t_{FF}\sum_{<ij>}\eta_{i}\eta_{j}cos(\theta_{ij}/2)(\tilde{f}_{i}^{\dagger}\tilde{f}_{j}) \cr
&&
+\tilde{\epsilon_{Fe}}\sum_{i}\eta_{i}\tilde{f}_{i}^{\dagger}\tilde{f}_{i}+
\epsilon_{Mo}\sum_{i\sigma}(1-\eta_{i})m_{i\sigma}^{\dagger}m_{i\sigma} \cr
&&+J_{AF}\sum_{<ij>}\eta_{i}\eta_{j}S_{i}\cdot S_{j} 
\label{Jinfinityham}
\end{eqnarray}

\begin{figure*}
\includegraphics[width=4.5cm,height=4.5cm,angle=-90,clip=true]{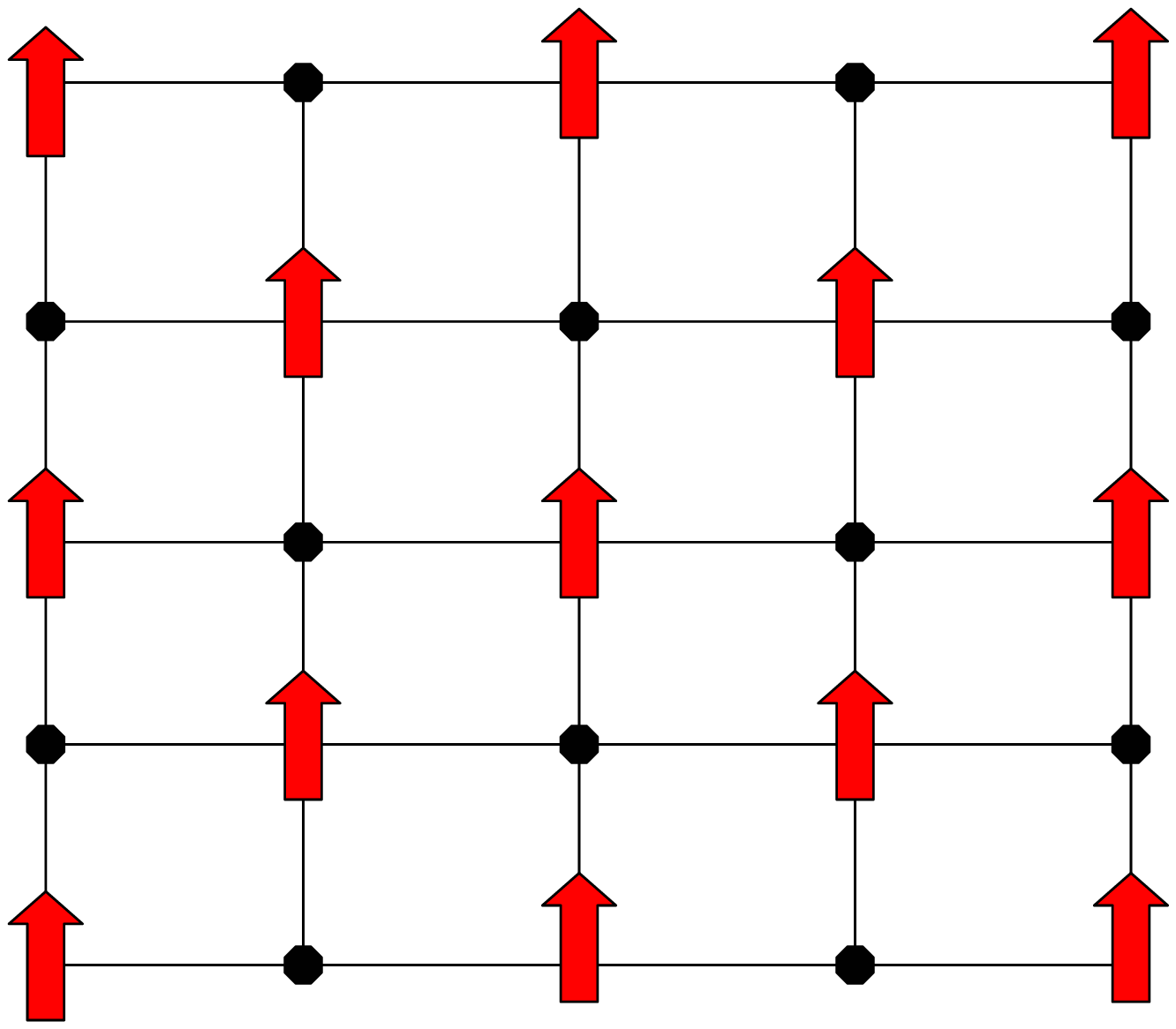}
\hspace{.9cm}
\includegraphics[width=4.5cm,height=4.5cm,angle=-90,clip=true]{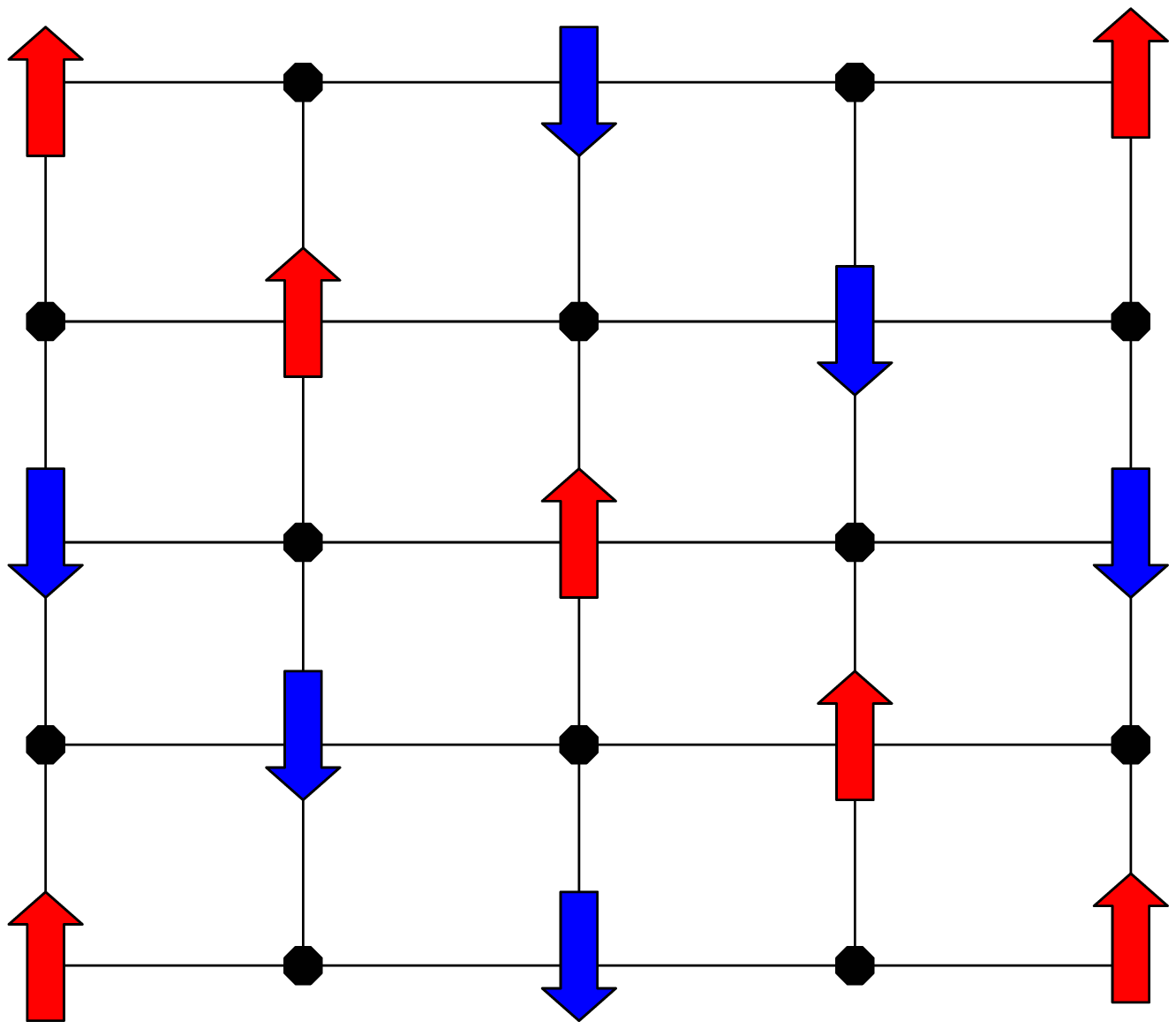}
\hspace{.9cm}
\includegraphics[width=4.5cm,height=4.5cm,angle=-90,clip=true]{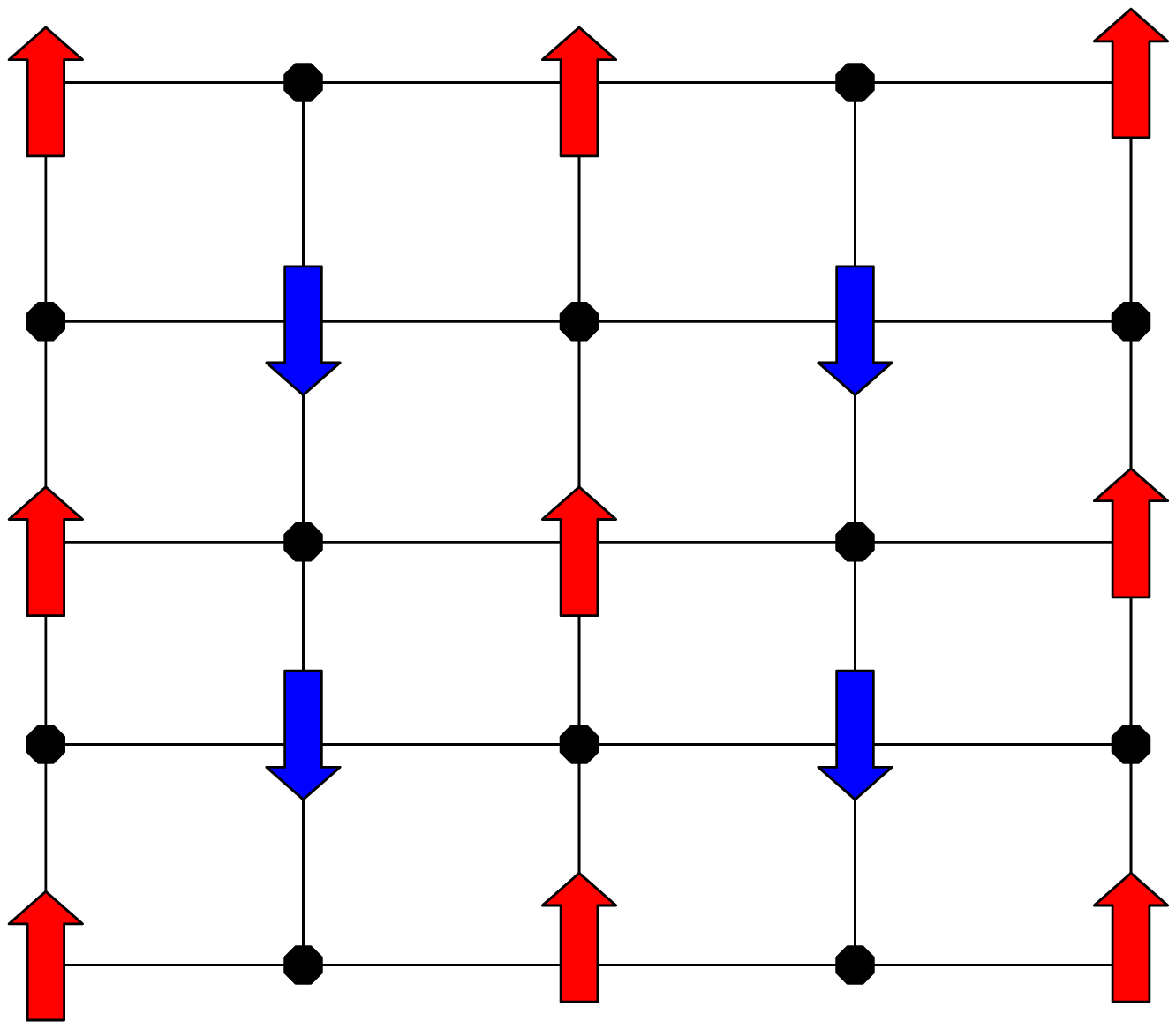}
\caption{The three magnetic phases in the ordered 2D model. Arrows represent Fe sites, and dots Mo sites.\\
Left: ferromagnetic (FM), center: 
antiferromagnet (AFM1), right:  antiferromagnet 
(AFM2). The moments are on the B sites.
}
\label{fig: 3phases}
\end{figure*}

\begin{figure*}
\includegraphics[width=5cm,height=6cm,angle=0,clip=true]{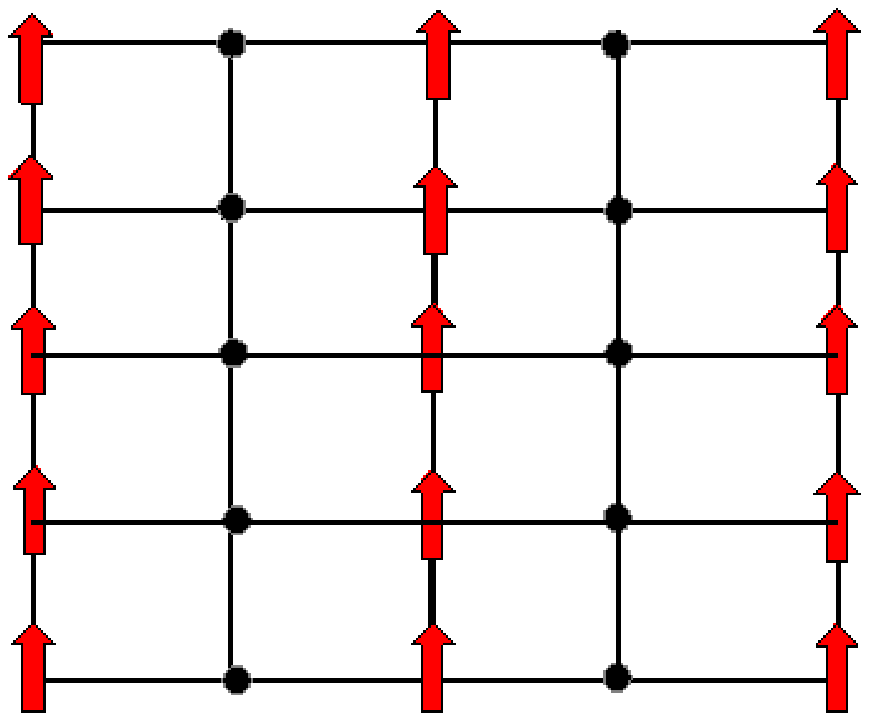}
\hspace{.9cm}
\includegraphics[width=5cm,height=6cm,angle=0,clip=true]{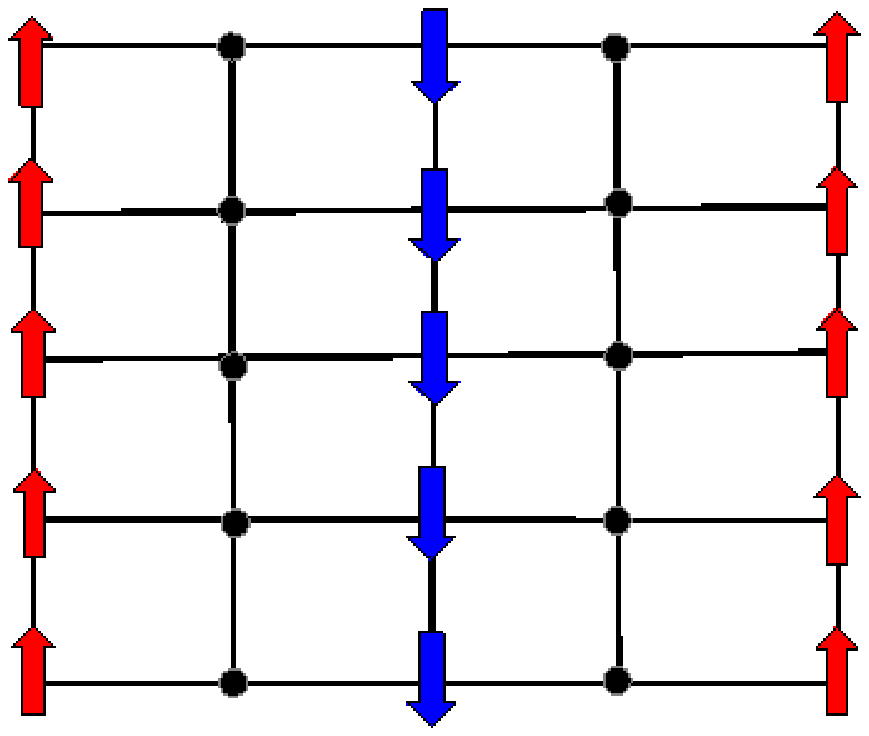}
\hspace{.9cm}
\includegraphics[width=5cm,height=6cm,angle=0,clip=true]{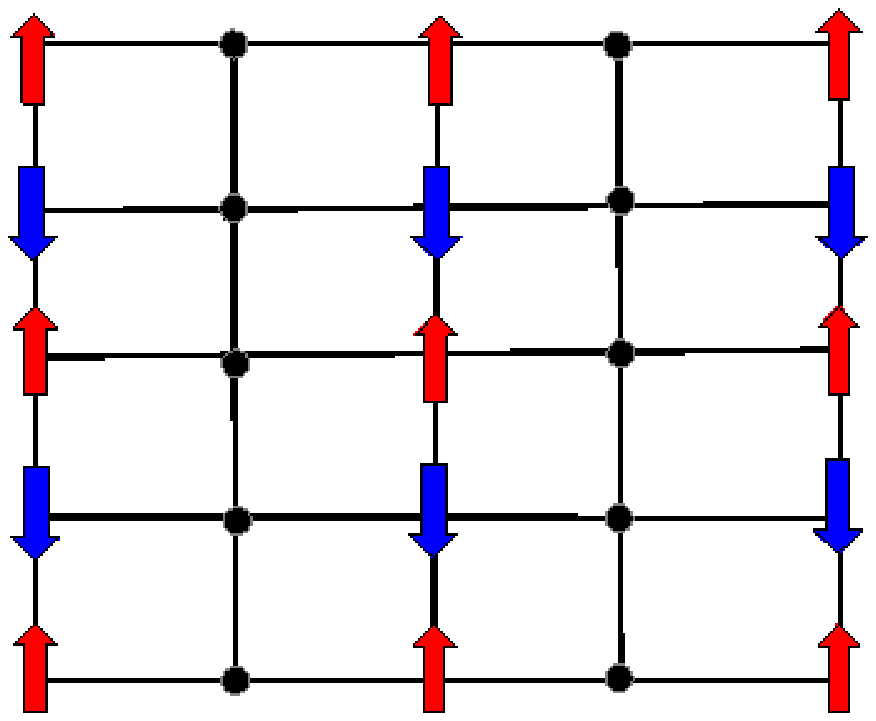}
\caption{The three magnetic phases in the  phase segregated case: alternate Fe and Mo-rich regions. Arrows represent
Fe sites, and the dots Mo sites.\\
Left: ferromagnetic (FM), center: 
antiferromagnet (AFM1), right:  antiferromagnet 
(AFM2). 
}
\label{fig: antisite_FM_AFMA_AFMG}
\end{figure*}

In the ordered case, this model has been studied as a function of filling of the Fe and Mo degrees of freedom, using various analytical and numerical methods~\cite{mePinaki}. It was found that while a ferromagnetic phase was 
stable for low fillings $n<1$, it became unstable to antiferromagnetic phases for larger filling $n>1$. There 
appeared to be two main types of antiferromagnetic phases: A type ($0-\pi$) or AFM1, and G type ($\pi-\pi$) phases, or AFM2. A schematic of these three phases are shown in Fig~\ref{fig: 3phases}.

\section{Alternate Phase segregated case}
              
Samples of La-overdoped SFMO are found to order in a patchy structure~\cite{Sugatapriv} where phase segregation 
occurs into alternate Fe-rich and Mo-rich regions. This leads to a substantial amount of antisite defects. The
simplest model for this situation can be considered in 2D as follows. Let us consider strips of Fe and Mo atoms
 arranged alternately as shown in Fig~\ref{fig: antisite_FM_AFMA_AFMG}, 
 or in the bottom panel of Fig~\ref{domains}. 
This would mean alternate lines of antisite defects separated from each other
 by patches where Fe and Mo are nearest neighbour. Such a patchy structure can be used to effectively describe
 the samples of Sr$_{2-x}$La$_{x}$FeMoO$_{6}$. This configuration can be thought to be realized from an ordered
 checkerboard configuration of Fe, Mo by proliferation of domain line-defects of Fe and Mo combined. Accordingly,
  the ferromagnetic, A type antiferromagnet and G type antiferromagnet can be realized in the the form shown in
 Fig~\ref{fig: antisite_FM_AFMA_AFMG}. In this case, the dispersions in the three phases can be calculated as follows.
  In the fully ferromagnetic case, the dispersion is given by:
\begin{widetext}
\begin{eqnarray}
\epsilon_{k}&=&\frac{[\epsilon_{Fe}+\epsilon_{Mo}+2(t_{FF}+t_{MM})cosky]\pm 
 \sqrt{[(\epsilon_{Fe}+2t_{FF}cosky)-(\epsilon_{Mo}+2t_{MM}cosky)]^{2}+16t_{FM}^{2}cos^{2}kx}}{2}   
\end{eqnarray}  
\end{widetext} 

If we put $t_{FF}=t_{MM}=0$, then 
\begin{eqnarray}
\epsilon_{k}=\frac{(\epsilon_{Fe}+\epsilon_{Mo})\pm \sqrt{(\epsilon_{Fe}-\epsilon_{Mo})^{2}+16t_{FM}^{2}cos^{2}kx}}{2}
\end{eqnarray}
which is just the band structure of a 1D binary lattice, appropriate for the isolated Fe-Mo alternating chains in
 the x-direction (to the right). If $\epsilon_{Fe}=\epsilon_{Mo}=0$, then 
$\epsilon_{k}=\pm2t_{FM}coskx$, revealing the 1D character. If we put $t_{FM}=0$, then we get, for $\epsilon_{Fe}=\epsilon_{Mo}=0$, $\epsilon_{k}^{+}=2t_{FF}cosky$ and $\epsilon_{k}^{-}=2t_{MM}cosky$, 
once again 1D band structures, this time in the y direction, as expected.

 For the antiferromagnetic cases, the dispersions are obtained in the following manner.
For the  A-type antiferromagnet, there are alternate ferromagnetic chains arranged antiferromagnetically. This
 can be obtained in the phase segregated limit by considering the fact that in the ordered phase (as shown in
 the central panel of Fig~\ref{fig: 3phases}), the up and
 down spins alternate on the Fe sites on any line along the x-direction, separated by an Mo site. 
 Since this structure is preserved by a Fe-Mo line domain shift in the x-direction,
  the resulting phase is as shown in the central panel of 
 Fig~\ref{fig: antisite_FM_AFMA_AFMG}.
 The dispersion is given by  $\epsilon_{k}=\epsilon_{Mo}+2t_{MM}cosky$ and 
\begin{widetext}
\begin{equation}
\epsilon_{k}=\frac{[\epsilon_{Fe}+\epsilon_{Mo}+2(t_{FF}+t_{MM})cosky]\pm\sqrt{[(\epsilon_{Fe}+2t_{FF}cosky)-
(\epsilon_{Mo}+2t_{MM}cosky)]^{2}+8t_{FM}^{2}}}{2}
\end{equation}
\end{widetext}


If we consider $t_{FF}=t_{MM}=0$, and $\epsilon_{Fe}=\epsilon_{Mo}=0$, then the dispersion simplifies to just
 three possible values: $0,\pm\sqrt{2}t_{FM}$ appropriate for a triatomic molecule, as expected for an Mo-Fe-Mo 
trimer. If $t_{FM}=0$,
then it gives the 1D dispersions $\epsilon_{Fe}+2t_{FF}cosky$ and $\epsilon_{Mo}+2t_{MM}cosky$, 
corresponding to 1D Fe chains and Mo chains in the y-direction as expected. 

The G-type antiferromagnet, which corresponds to the $\pi-\pi$ antiferromagnet for the ordered lattice, can be
 realized in this antisite disordered case by considering the fact 
 that along a line in the x-direction, in the ordered phase, the
 Iron spins all remain parallel (see right panel of Fig~\ref{fig: 3phases}). 
 Since this is once again preserved during a Fe-Mo line domain shift in the x-direction, 
 hence the spin configuration that is realized is as shown in the right panel of 
 Fig~\ref{fig: antisite_FM_AFMA_AFMG}. Here the Iron spins
 are parallel along the x-direction, while they alternate antiparallely along the y-direction. In this case,
  the dispersion is obtained as the roots of a cubic equation:
\begin{widetext}
\begin{eqnarray}
(\epsilon_{Fe}-\epsilon_{k})(\epsilon_{Mo}-\epsilon_{k})^{2}-(\epsilon_{Fe}-\epsilon_{k})t_{MM}^{2}(2+2cosky)
-t_{FM}^{2}(2+2coskx)(\epsilon_{Mo}-\epsilon_{k})=0
\end{eqnarray}
\end{widetext}
The limits can be obtained exactly. For example if $t_{MM}=0$, $\epsilon_{k}=\epsilon_{Mo}$ 
(the isolated Mo level) and 
\begin{eqnarray}
\epsilon_{k}=\frac{(\epsilon_{Fe}+\epsilon_{Mo})\pm\sqrt{(\epsilon_{Fe}-\epsilon_{Mo})^{2}+16t_{FM}^{2}cos^{2}kx/2}}{2}
\end{eqnarray}
If we put $\epsilon_{Fe}=\epsilon_{Mo}=0$, this emerges as a 1D band structure
 $\epsilon_{k}=\pm 2t_{FM}coskx/2$, as the structure reduces to isolated
 binary Fe-Mo  chains in the x-direction.
If, on the other hand, we put $t_{FM}=0$, then the band structure reduces to 
 $\epsilon_{k}=\epsilon_{Fe}$, the isolated Fe level, and
 the 1D bands $\epsilon_{k}=\epsilon_{Mo}\pm 2t_{MM}cosky/2$ in the y-direction, as expected.

\section{Results}

\begin{figure}
\includegraphics[width=8cm]{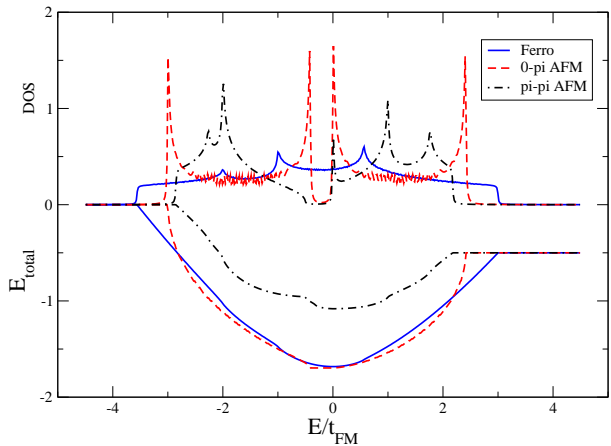}
\caption{\label{DOS_E} DOS for the three phases (top) and total energy of the three phases (bottom) plotted vs. energy/chemical potential} 
\end{figure}

The DOS for the three phases (ferro, A-type AFM, G-type AFM) as a function of energy are presented in Fig~\ref{DOS_E}, on the upper side of the y-axis. On the lower side, the total energy of the three phases 
  are given as a function of energy, now to be interpreted as chemical potential $\mu$. 
  It is observed that the bandwidth of the ferro phase is the maximum, while
 that of the two antiferro phases are close to each other, although that of the A-type phase is slightly
 larger. Both the antiferromagnetic phases have Van Hove 
 singularities revealing their lower dimensional character. 
The energy of the ferro phase is lowest for small values of filling simply 
because of the large bandwidth of the ferro phase. This means that the ferro phase would be most stable for 
low electron filling. However, as the filling increases, for large regions of $\mu$, the A-type
antiferromagnetic phase is energetically more stable than the ferro phase. Since this is a kinetic energy driven
 stabilization, hence this means that an antiferromagnetic metal phase would be stabilized for large regions of
 filling. Except of course the small portion of chemical potential where there is a gap in the DOS, arising from
 the charge transfer energy $\Delta=\epsilon_{Fe}-\epsilon_{Mo}$, which we have deliberately 
 considered to be finite. This small portion of filling would represent the only insulating behaviour. 
 The dominant region of filling, for large enough filling, continues to be antiferromagnetic metallic, thereby
 underlining that the same kinetic energy driven mechanism which works for the ordered phase 
 continues to stabilize this AFM phase even in the presence of antisite defects, in a phase segregated scenario.

\begin{figure}
\includegraphics[width=5cm,height=5cm,angle=0,clip=true]{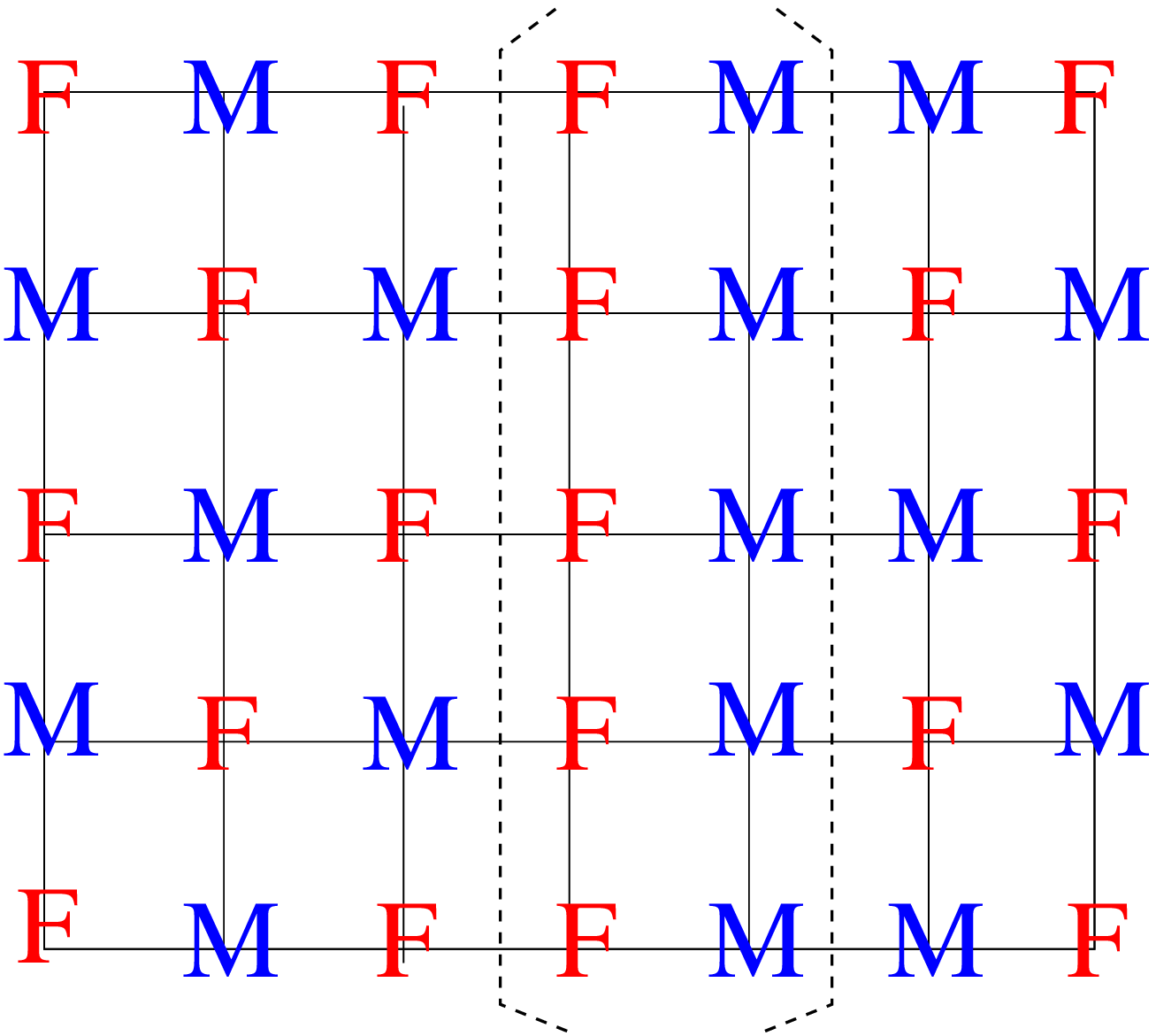}
\vspace{.5cm}
\includegraphics[width=5cm,height=5cm,angle=0,clip=true]{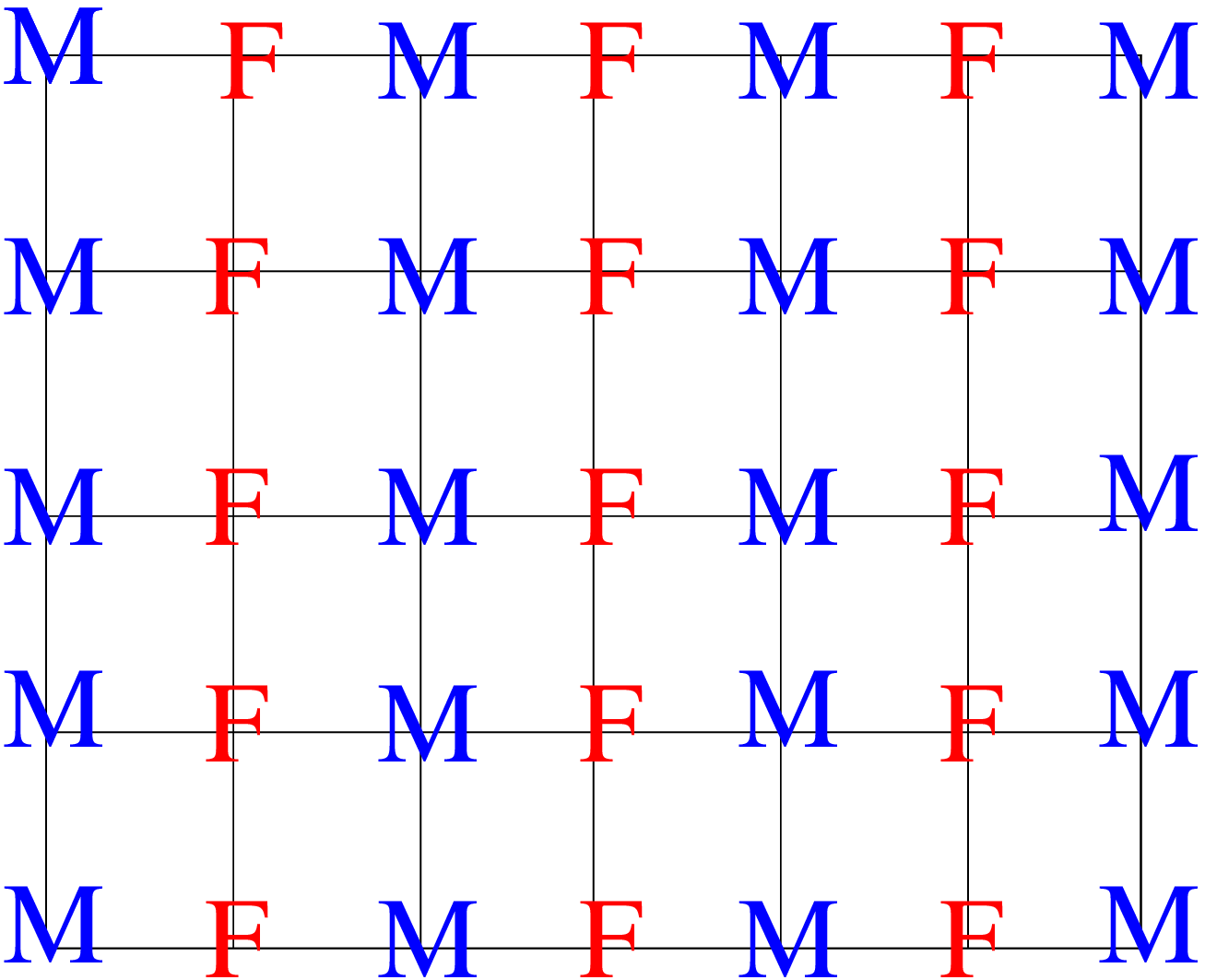}
\caption{ Top: A single domain line (combined Fe-Mo) antisite region in an otherwise ordered sample. F-srepresent the Iron atoms,and M-s the Molybdenum atoms. Bottom: Proliferation of such domains can result in the alternate Fe-Mo rich antisite structure depicted in Fig~\ref{fig: antisite_FM_AFMA_AFMG}.}
\label{domains} 
\end{figure}


Interestingly, though, the remnant of the G-type antiferromagnet, most stable for large filling in the ordered
 phase~\cite{mePinaki}, is not found to be stabilized kinetically in any part of the filling in this phase 
 segregated case. 


\begin{figure}
\includegraphics[width=8cm]{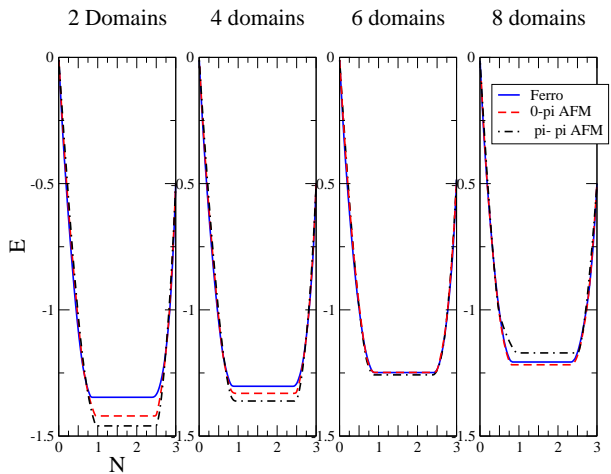}
\caption{ Energy vs filling for the three phases, compared for configurations with 2,4,6 and 8
 domain Fe-Mo antisite defects.} 
\label{domains_ED}
\end{figure}

 In order to investigate how the kinetic energy driven stabilization behaviour changes from the ordered phase
 to the phase segregated phase with antisite defects, we have also considered intermediate phases with a combined
 Fe-Mo domain structure which is linear in the y direction (Fig~\ref{domains}).
 Using Exact Diagonalization calculations in real
 space, we have obtained the energies of the three phases, variationally, as a function of filling. 
 Parameters used are $t_{FM}=1$,$t_{MM}=0.75$,$t_{FF}=0.5$ and $\Delta=0.5$. 
 We considered intermediate configurations with 2,4, 6 and 8 domains in a $16\times16$ lattice. This symbolizes
 propagation of the linear domains in the x direction. It is observed (Fig~\ref{domains_ED}) 
 that the antiferromagnetic phases slowly
 interchange their stability behaviour as the domains propagate through the lattice. There is a smooth
 transition from stabilization of the G-type phase in the low domain case to the stabilization of the 
 A-type phase in the case with large number of domains. Thus a smooth interpolation is obtained from the ordered
 phase to the phase segregated phase, proving the scenario proposed earlier, and showing the continued
 kinetic energy induced stabilization of an antiferromagnetic metal phase for large filling; 
only the character of the antiferromagnetic phase changes.  
 Although the G type phase is no longer stabilized kinetically, the superexchange, being nearest
 neighbour for the phase segregated case, and antiferromagnetic, can stabilize it, and bring its energy close
 to that of the A-type phase, even possibly surpassing it at some filling~\cite{superex}. Such a scenario can however, occur only for a small region of filling, as the G-type phase has the lowest bandwidth. Even then, the ground state continues to be an antiferromagnetic metal at high filling;
 either A type or G type, depending in the filling.
  It is interesting to note that even from 
 kinetic energy considerations, the energy difference between the ferro and the ground state antiferromagnetic
 phase is much larger for the ordered case rather than in the phase segregated limit, as can be understood
 by comparing the figures for 2 domains and 8 domains. 
  This can provide a plausible explanation to the signature of phase coexistence observed
 in the actual experiment using the phase segregated samples.~\cite{Sugatapriv}.

\section{Summary and Outlook} 

 We have considered the simplest model of phase segregation of B-site (Fe) and B$\prime$ site (Mo) 
 in double perovskite Sr$_{2}$FeMoO$_{6}$, leading to antisite regions. Using variational methods, 
 we have shown that an antiferromagnetic
 metal phase can still be stabilized in large regions of filling. The nature of the antiferromagnetic phase
 that is kinetically stabilized, is however, found to change as the system goes from ordered to phase segregated.
 Using extensive simulations in the intermediate phases involving antisite domains, 
 we have verified that this transition is smooth,
  and the system remains an
 antiferromagnetic metal throughout for large regions of the filling, only the nature of the antiferromagnet
 changes from the ordered to the phase segregated limit.

\section{Acknowledgment}
The author gratefully acknowledges discussions with   P. Majumdar, S.ray  and T. Saha Dasgupta.


\begin{thebibliography}{99}
\bibitem{superex} The nearest neighbour superexchange (active in antisite regions)
 affects the ferromagnetic phase
  and the A-type antiferromagnetic phase in the same way, and hence does not change their relative energies. 
 Only the G type phase gets affected differently.
\end{thebibliography}
\end{document}